# Unraveling the Biomarker Prospects of High-Altitude Diseases: Insights from Biomolecular Event Network Constructed using Text Mining

Balu Bhasuran[1,2], Sabenabanu Abdulkadhar[3,4], Jeyakumar Natarajan[1,4*]

[1]DRDO-BU Center for Life Sciences, Bharathiar University Campus, Coimbatore, Tamil Nadu 64104, India.

[2]School of Information, Florida State University, Tallahassee, Florida 30306, USA

[3]Department of CSE, Koneru Lakshmaiah Education Foundation, Vaddeswaram, Andhra Pradesh 522302, India.

[4]Data Mining and Text Mining Laboratory, Department of Bioinformatics, Bharathiar University, Coimbatore, Tamil Nadu  641046, India.

**\*Correspondence should be addressed to Jeyakumar Natarajan;**
n.jeyakumar@yahoo.co.in

## Abstract

High-altitude diseases (HAD), encompassing acute mountain sickness (AMS), high-altitude cerebral edema (HACE), and high-altitude pulmonary edema (HAPE), are triggered by hypobaric hypoxia at elevations above 2,500 meters. These conditions pose significant health risks, yet the molecular mechanisms remain insufficiently understood. In this study, we developed a biomolecular event extraction pipeline integrating supervised machine learning with feature-based and multiscale Laplacian graph kernels to analyze 7,847 curated HAD-related abstracts from PubMed. We extracted over 150 unique biomolecular events—including gene expression, regulation, binding, and localization—and constructed a weighted, undirected biomolecular event network comprising 97 nodes and 153 edges. Using the PageRank algorithm, we prioritized key biomolecules based on their centrality within the event network. The top-ranked proteins included Erythropoietin (EPO) (0.0163), Vascular endothelial growth factor (VEGF) (0.0148), Hypoxia-inducible factor 1 (HIF-1) alpha  (0.0136), Endothelial PAS Domain Protein 1 (EPAS1) and Angiotensin-Converting Enzyme (ACE) (0.0119), Egl nine homolog 1 (EGLN1), Endothelin 1 (ET-1), and 70 kilodalton heat shock protein (Hsp70)(0.0118), all of which play crucial roles in oxygen sensing, vascular remodeling, erythropoiesis, and blood pressure regulation. Subnetwork analysis revealed three major functional clusters centered on hypoxia response, inflammation, and stress adaptation pathways. Our integrative approach demonstrates the utility of large-scale text mining and graph-based analysis to uncover mechanistic insights and prioritize potential biomarkers for high-altitude disease.

## Introduction

Altitude sickness, also known as high-altitude illness, is a physiological response to reduced oxygen availability at elevations typically above 2,500 meters. It manifests in three major clinical forms: acute mountain sickness (AMS), high-altitude cerebral edema (HACE), and high-altitude pulmonary edema (HAPE) [1,2]. Common symptoms include headache, nausea, loss of appetite, and sleep disturbances [3]. The risk increases with rapid ascent, and millions of individuals—ranging from travelers and athletes to military personnel—are exposed to such environments each year. The onset of high-altitude diseases (HAD) can occur suddenly and may range from mild discomfort to severe, life-threatening conditions [4].

Despite extensive research efforts, much of the valuable biomedical knowledge generated from experimental studies remains locked within the vast and rapidly growing body of scientific literature. Biomedical text mining has emerged as a powerful tool to bridge this knowledge gap by uncovering hidden patterns, generating hypotheses, and identifying novel associations from unstructured text [5–10]. Applications of text mining have shown promise in various domains such as gene-disease associations, drug repurposing, and pharmacogenomics [11–15]. In particular, understanding how genes and proteins behave under hypoxic stress—through events like transcription, regulation, expression, localization, and binding—offers critical insight into the pathophysiology of HAD [16,17]. These biomolecular events play a central role in functional genomics, disease mechanisms, and therapeutic targeting [18]. Automatically extracting such events into structured representations, a task known as biomolecular event extraction [19–25], remains technically challenging due to issues such as inconsistent terminology, gene/protein name ambiguity, and varying event descriptions [26,27].

To address these challenges, we developed a supervised machine learning-based text mining system for extracting biomolecular events associated with HAD. Using a curated corpus of PubMed abstracts, our system identified key genes, proteins, and their interactions, and constructed a biomolecular interaction network. We then applied the PageRank algorithm to prioritize proteins based on their centrality and connectivity within the network. Our analysis revealed recurring molecular events and highlighted candidate biomarkers potentially involved in the progression of HAD. The remainder of this paper is organized as follows: the Materials and Methods section details the corpus construction and event extraction methodology; the Results section presents the identified biomolecular events and network characteristics; the Discussion interprets key molecular findings; and finally, we conclude with future research directions.

## Materials and Methods

The biomolecular event extraction pipeline for High-Altitude Disease (HAD) literature was designed as a multi-step process comprising: data collection, preprocessing, event extraction, network construction, network analysis, and biomarker discovery. For data collection, we utilized PubMed (https://pubmed.ncbi.nlm.nih.gov/), a widely used search engine for accessing the MEDLINE database maintained by the National Center for Biotechnology Information (NCBI), U.S. National Library of Medicine (NLM). Relevant HAD literature was retrieved using targeted keyword queries. The resulting records were formatted by extracting the PubMed ID (PMID), title, and abstract for each entry. Preprocessing involved several natural language processing tasks, including sentence splitting, tokenization, part-of-speech (POS) tagging, chunking, and biomedical lemmatization. Syntactic information was then extracted using dependency and phrase structure parsing, followed by the extraction of 14 domain-specific linguistic and semantic features aimed at capturing biomolecular event mentions.

To perform event extraction, we employed a hybrid machine learning approach that integrates a Multiscale Laplacian Graph (MLG) kernel with a traditional linear feature-based kernel within an ensemble learning framework. This model was used to detect event triggers, while a rule-based pattern matching engine (using regular expression templates) was implemented to identify event arguments such as themes and causes. The extracted genes/proteins and associated events were used to construct a biomolecular interaction network. We applied graph analysis—specifically the PageRank algorithm—to this network to identify and prioritize key proteins with potential biomarker roles in hypoxia-related pathways. The complete event extraction pipeline is illustrated in Figure 1 and is further elaborated in the sections that follow.

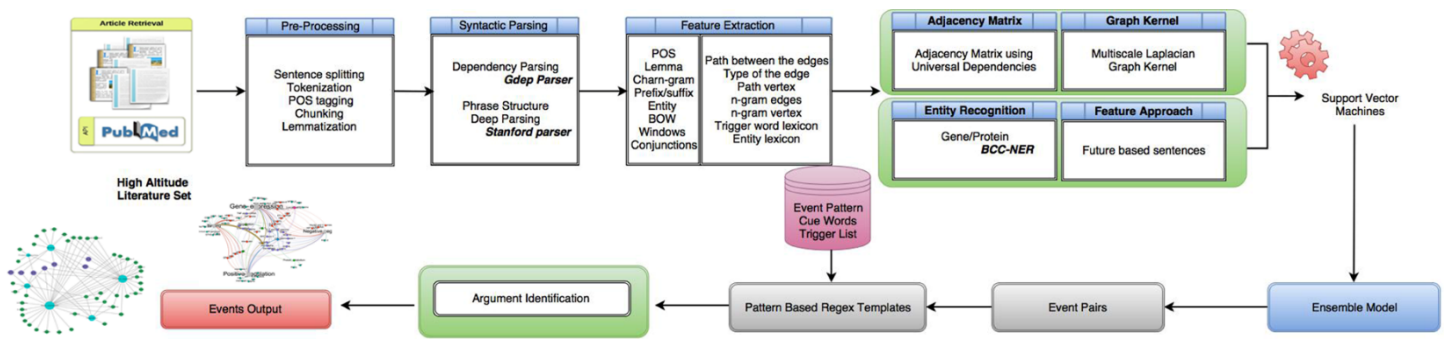

**Figure 1** Schematic architecture of the bi-molecular event extraction methodology for High Altitude Diseases

**High Altitude Disease Information Retrieval**

In the context of text mining, Information Retrieval (IR) refers to the process of gathering relevant literature on a specific topic from various biomedical databases. PubMed, a widely used search engine for accessing the MEDLINE bibliographic database—which contains over 34 million citations in the biomedical and life sciences domains—was utilized for this purpose. To compile a comprehensive corpus focused on High-Altitude Diseases (HAD), we queried PubMed using three key terms: *"High Altitude Diseases," "High Altitude Sickness,"* and *"High Altitude Illness."* The retrieved articles were downloaded and processed to create a curated HAD-specific text corpus, which included the PubMed ID (PMID), title, and abstract of each article. This corpus served as the foundation for subsequent information extraction and analysis.

**Biomolecular Event Extraction System**

Biomolecular events represent interactions and functional relationships among biological entities such as genes, proteins, and molecules. These events describe key processes—including transcription, regulation, phosphorylation, and binding—that drive cellular and physiological functions. Automatically extracting such events from unstructured biomedical literature and converting them into structured representations is known as event extraction. The resulting event structures are analogous to reactions in biological pathways and are valuable for applications such as pathway curation, biological network construction, and drug repurposing.

In this study, we applied our in-house developed biomolecular event extraction system [28] to a curated corpus on high-altitude diseases. The system operates in three primary stages: (1) preprocessing and cleaning of textual data, (2) event detection using a hybrid machine learning framework, and (3) event argument identification using pattern-matching techniques. Each step is described in detail below.

**Preprocessing and Cleaning**

The preprocessing phase involves preparing raw biomedical text for downstream event extraction. This includes sentence segmentation, tokenization, part-of-speech (POS) tagging, lemmatization, chunking, dependency parsing, and named entity recognition (NER). We used the OpenNLP toolkit [29] for sentence segmentation, POS tagging, and chunking, and BioLemmatizer [30] for lemmatization. Dependency parsing was carried out using both the Stanford Parser [31] and GDep Parser [32] to identify grammatical structures and syntactic dependencies between trigger words and biological entities. Accurate identification of named entities such as genes and proteins is critical for capturing event participants. For this, we employed BCC-NER [33], our in-house hybrid named entity recognition system, which automatically tags biomedical entities relevant to HAD.

**Ensemble Classifier**

To detect biomolecular events, we implemented a hybrid ensemble learning approach that combines a feature-based kernel with a Multiscale Laplacian Graph (MLG) kernel [34]. The feature-based kernel captures local textual patterns using token-level, sentence-level, and lexical features, while the MLG kernel models complex sentence structures and multilevel topological relationships among event nodes. These two kernels were integrated into an ensemble Support Vector Machine (SVM) framework [35], allowing the system to leverage both local and global context for robust event identification. Ensemble methods are increasingly popular in biomedical text mining due to their improved generalizability and high classification accuracy [36,37].

**Event Argument Detection**

The final step involves identifying the roles of entities involved in each event, such as THEME and CAUSE. We employed a lexico-syntactic pattern-based approach, enhanced with semantic role labeling, to extract arguments accurately from complex biomedical sentences. Lexico-syntactic patterns represent generalized linguistic templates that link trigger words (e.g., *activate*, *bind*) with associated biological entities. These patterns were derived through distributional analysis of the training data and refined into rule-based templates for argument detection. Manual rule crafting ensured precise identification of event participants, thereby improving the quality of the extracted event structures. This final structured output was used for downstream network construction and biomarker prioritization.

**Biomolecular Event Network**

The extracted genes, proteins, and associated biomolecular events were integrated to construct a biomolecular event network specific to High-Altitude Diseases (HAD). For network construction and visualization, we utilized Gephi version 0.9.2 [38], a widely used platform for interactive graph analysis. The resulting network was modeled as an undirected, weighted graph, where biomolecular events (e.g., gene expression, regulation) functioned as central hub nodes, and edges represented co-occurrences of genes/proteins within those events. Edge weights were assigned based on the frequency of co-occurrence in the HAD corpus, allowing for the representation of interaction strength.

To identify the most influential biomolecules within this network, we applied the PageRank algorithm [39], a node-ranking technique originally developed by Google for ranking web pages. PageRank evaluates the importance of nodes based on both the number and quality of connections they receive. In our context, this allowed us to compute key network attributes—such as PageRank score, weighted in-degree, weighted out-degree, and min-max normalized rank—for each node. These attributes were used to identify critical proteins and to generate subnetworks that highlight the most biologically relevant interactions.

PageRank has been successfully applied in previous biomedical studies for ranking concepts in association networks, identifying research subfields, and re-ranking scientific literature [40–45]. In this study, the algorithm was used to quantify the centrality of each gene/protein in relation to frequently co-occurring events. Nodes with higher PageRank scores were interpreted as more functionally significant in the context of HAD pathogenesis. The network thus becomes a structured map of biomolecular interactions, where hubs (events) and their linked proteins provide insights into biological mechanisms. Inspired by earlier applications, we used PageRank not only to prioritize biomarker candidates but also to construct a normalized subnetwork for deeper analysis of gene-event relationships relevant to high-altitude adaptation and disease. The implementation of PageRank is explained below.

$$pr(u) = c \sum_{v \in b(u)} \frac{pr(v)}{N_v} \qquad (1)$$

Here we have taken $u$ is a node in the network. $b(u)$ is a set of nodes that represent to node $u$, $pr(u)$ and $pr(v)$ are rank scores of the nodes $u$ and $v$ respectively. $N_v$ denotes number of outgoing edges of the node $v$. $c$ is the factor used to normalize the nodes in the network. Next we have assigned the weights to each node in the network. Then, we measured the weights of inner edges $W_{(v,u)}^{in}$ and outer edges $W_{(v,u)}^{out}$ of the nodes $u$ and $v$. $w_{(v,u)}^{in}$ is a weight of edge $(v,u)$ calculated based on the number of inner edges of node u and the number of inner edges of v. $w_{(v,u)}^{out}$ is the weight of edge $(v,u)$ calculated based on the number of outer edges of node u and number of outer edges of v. Finally, we employ PageRank to weighted nodes in the network.

$$pr(u) = (1-d) + d \sum_{v \in b(u)} pr(v)\, W_{(v,u)}^{in}\, W_{(v,u)}^{out} \qquad (2)$$

Here $w$ is the dampening factor which was used to calculate the probability of edges of each node in the network and $(1-d)$ is the page rank distribution from non-directly connected nodes.

**Results**

Text mining has proven to be a powerful tool in biomedical research, enabling the discovery of novel biomarkers, uncovering implicit relationships, and generating informative literature-derived networks. These approaches often provide fresh insights into the underlying biological mechanisms of specific diseases [12, 46–48]. In our earlier work, we applied literature mining techniques to investigate gene–gene associations in the context of high-altitude diseases, revealing functional relationships and shedding light on key genetic interactions involved in hypoxia adaptation [12]. Building on that foundation, the present study focuses on extracting biomolecular events related to high-altitude diseases using advanced text mining methodologies for the purpose of biomarker discovery. The following results outline the corpus generation process and the findings from our event extraction and network analysis.

**High Altitude Disease Corpus Generation**

To construct a comprehensive dataset for event extraction, we performed a targeted keyword search in PubMed up to the year 2024, using three synonymous terms: *High Altitude Disease*, *High Altitude Sickness*, and *High Altitude Illness*. The search results were downloaded in MEDLINE format as three separate text files. These files were merged and normalized by removing duplicate records, resulting in a non-redundant dataset. Each entry was formatted to include the PubMed ID (PMID), title, and abstract, forming the basis of the HAD corpus. Manual curation was conducted to ensure that all retained articles were directly relevant to high-altitude disease research, encompassing both primary studies and reviews. The number of articles retrieved for each keyword query is summarized in Table 1, which forms the foundation for all subsequent biomolecular event extraction and analysis.

Table 1 High Altitude Disease dataset downloaded from PubMed using keyword query

| Keyword | Query | PubMed Output |
|---|---|---|
| High Altitude Diseases | | 5422 records |
| High Altitude Sickness | | 3904 records |
| High Altitude Illness | | 1023 records |
| **Total** | | **10349 records** |

A retrieved HAD dataset consisted of 10349 records, among which 7847 records were used for advanced investigation, subsequent to removing the covering overlapping PubMed IDs and non-relevant documents to the study.

**High Altitude Disease Event Extraction**

Studies on HAD data revealed various biological processes involved with genes and proteins in the altitude sickness associated with most biomedical events like expression, regulation, positive, negative regulation, transcription, and binding, etc. The identified events can be used to automatically construct the pathway of altitude sickness.

The biological processes exposed that genes and proteins in altitude sickness are associated with most biomedical events like expression, regulation, positive, negative regulation, transcription, and binding, etc. The identified events were used to automatically construct the pathway of altitude sickness from biomedical literature. The following examples explained that the high-altitude disease corpus has numerous events, and these events were used to construct the pathway and identify the drug targets.

Phosphorylation played an important role in supplying the oxygen of the whole body. In the above example 1, proteins, AMP-activated protein kinase, A2B adenosine receptor, BPG mutase, and AMP-activated protein kinase enhanced BPG mutase activation were the most relevant proteins and highly participated in the event of Phosphorylation. Our ensemble model efficiently extracts the event Phosphorylation and associated proteins to construct the pathway and identify the drug targets in hypoxia.

**Example: 1**

27482003      Mechanistically, we demonstrated that erythrocyteAMP-activated protein kinase was activated in humans at high altitude and that AMP-activated protein kinase is a key protein functioning downstream of the A2B adenosine receptor, phosphorylating and activating BPG mutase and thus inducing 2,3-BPG production and O2 release from erythrocytes. Significantly, preclinical studies demonstrated that activation of AMP-activated protein kinase enhanced BPG mutase activation, 2,3-BPG production, and O2 release capacity in CD73-deficient mice, in erythrocyte-specific A2B adenosine receptor knockouts, and in wild-type mice and in turn reduced tissue hypoxia and inflammation.

*T0*    ***Protein*** *29 101*      *Erythrocyte Adenosine A2B Receptor-Mediated AMP-Activated Protein Kinase*
*T1*    ***Protein*** *365 387*      *A2B adenosine receptor*
*T2*    ***Protein*** *424 434*      *BPG mutase*
*T3*    ***Protein*** *589 656*      *AMP-activated protein kinase enhanced BPG mutase activation*
*T1038* ***Phosphorylation*** *390 405*
*T1039* ***Positive regulation*** *574 584*
*E116*   ***Phosphorylation:*** *T1038* ***Theme:*** *T0*
*E117*   ***Phosphorylation:*** *T1038* ***Theme:*** *T1*
*E118*   ***Phosphorylation:*** *T1038* ***Theme:*** *T2*
*E119*   ***Positive_regulation:*** *T1039* ***Cause:*** *T3* ***Theme:*** *T3*

In example 2, the proteins Angiotensin II Receptor 1 gene variants, Angiotensin II Receptor 1 (AGTR1), were incorporated with the event 'binding', and they played a key role in developing the altitude illness high-altitude pulmonary edema (HAPE).

**Example: 2**

27732943      Angiotensin II Receptor 1 gene variants are associated with high-altitude pulmonary edema risk. Previous studies demonstrated that Angiotensin II Receptor 1 (AGTR1) may playan important role in the development of high-altitude pulmonary edema.

*T0*    ***Protein*** *1 40*    *Angiotensin II Receptor 1 gene variants*
*T1*    ***Protein*** *141 166*      *Angiotensin II Receptor 1*
*T1036* ***Binding*** *48 58*
*E113*   ***Binding:*** *T1036* ***Theme:*** *T0* ***Theme2:*** *T1*
*E114*   ***Binding:*** *T1036* ***Theme:*** *T1* ***Theme2:*** *T0*

In the following example 3, the proteins EPAS1, EGLN 1 were broadly associated with the event binding and this will help to detect the pathway of AMS. Our ensemble model effectively identified the gene expression and proteins from the corpus by the implementation of MLG and features.

**Example: 3**

---

22595196  Direct sequencing was used to identify individualgenotypes for the three SNPs. Age was found to be significantly associated with the EPAS1 SNP in the CMS patients while heart rate (HR) and oxygen saturation level of hemoglobin (SaO(2)) were found to be significantly associated with the EGLN1 (rs480902) SNP in the Han patients with AMS.

---

T2  ***Protein** 913 923  EPAS1*
T3  ***Protein** 999 1009 hemoglobin*
T4  ***Protein** 1079 1084 EGLN1*
T1003 ***Binding** 889 899*
T1004 ***Binding** 1055 1065*
E5  ***Binding:** T1004 **Theme:** T3*
E4  ***Binding:** T1003 **Theme:** T2*

---

During the identification of proteins and events associated with High Altitude Diseases (HAD), we observed that a specific subset of proteins consistently appeared as both Theme and Cause across multiple instances. This dual role suggests that these proteins play a central role in the underlying biological pathways and may serve as potential biomarkers for HAD. Furthermore, we found that four primary types of events—Gene Expression, Binding, Regulation, and Localization—frequently co-occur with these proteins. This recurrence indicates that these associations have been substantiated across multiple studies, reinforcing their biological significance.

A comprehensive list of all identified biomolecular events, along with their corresponding PMIDs and abstracts, is provided in Supplementary Data 1. The list of identified gene/protein names along with their associated event theme IDs is available in Supplementary Data 2, and the catalog of biomolecular events annotated with Theme and Event IDs is included in Supplementary Data 3.

**High Altitude Disease Biomolecular Event Network**

The biomedical event extraction model captures both simple and complex natural language expressions, enabling a wide range of downstream text mining applications such as pathway reconstruction, semantic search, and biomarker discovery [16–28]. In this study, the extracted biomolecular events served as the foundation for constructing a comprehensive network representing the molecular landscape of high-altitude diseases (HAD). This biomolecular event network integrates key genes and proteins that frequently participate in events such as gene expression, regulation, and binding—highlighting their potential roles as biomarkers or therapeutic targets.

The network structure emphasizes the functional interplay between biomolecules and biological processes, with biomolecular events acting as central hub nodes connected to associated gene/protein entities. Figure 2 illustrates this network, offering a systems-level view of the molecular interactions that underpin the pathophysiology of HAD. The network serves not only as a visual representation of literature-derived knowledge but also as a basis for quantitative analysis, including node ranking and subnetwork identification.

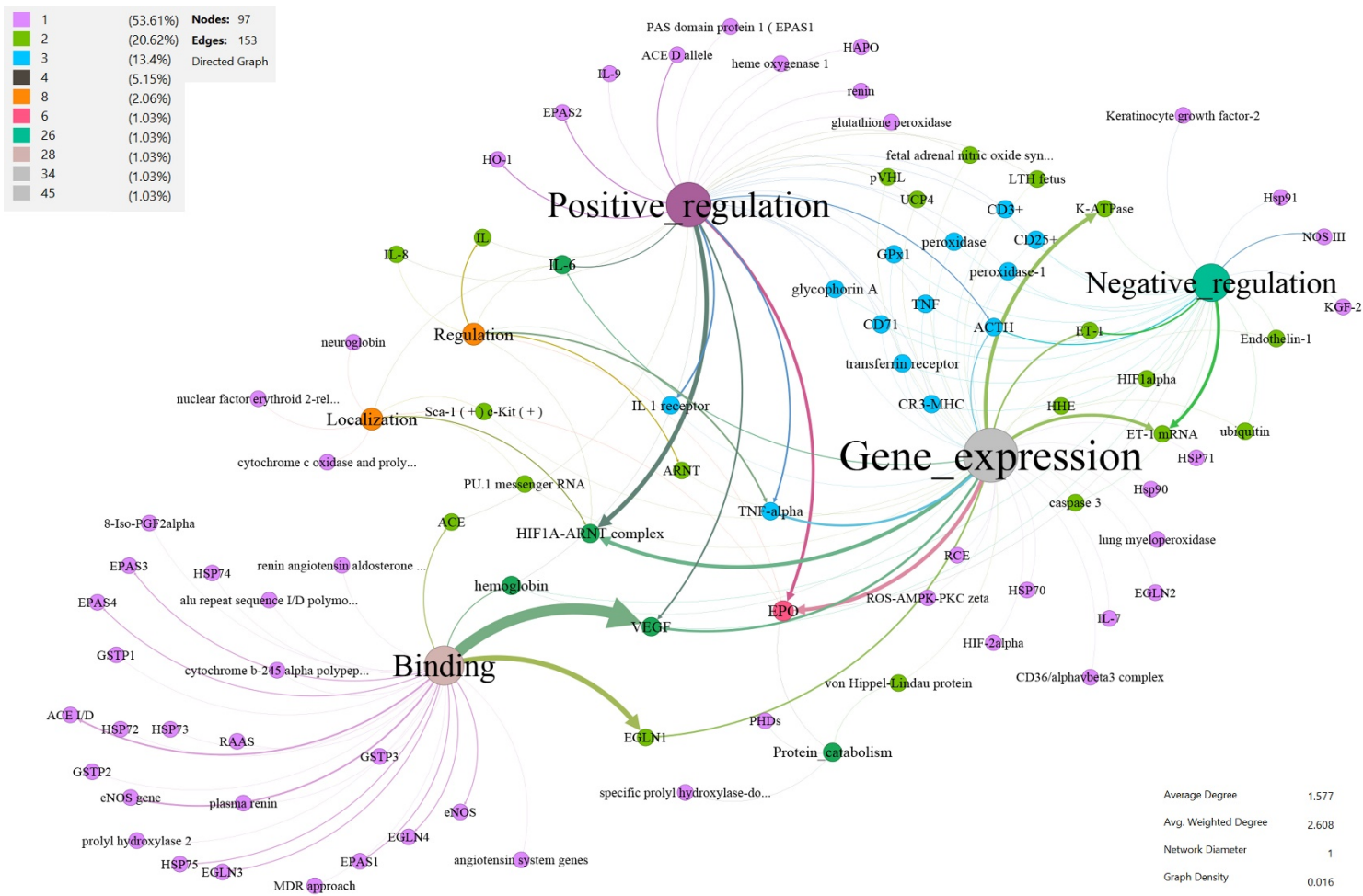

**Figure 2.** Biomolecular event network for high-altitude disease illustrating key molecular entities and biological processes involved in the response to hypoxic conditions.

A weighted, heterogeneous biomolecular event network was constructed using event themes extracted from the curated HAD literature. In this network, both genes/proteins and biomolecular events (e.g., gene expression, regulation) were represented as nodes. An edge was drawn between a gene/protein and an event node whenever the molecule functioned as a theme within that event. Each edge was assigned a weight proportional to its frequency of occurrence in the HAD corpus—higher co-occurrence frequencies resulted in stronger edge weights, reflecting greater biological relevance. The final network consisted of 97 nodes and 153 edges, capturing a broad spectrum of literature-derived molecular interactions related to high-altitude diseases. The network depicted in figure 2 revealed the major proteins which are acting in these events as Vascular endothelial growth factor (***VEGF***), Hypoxia-inducible factor 1(***HIF-1***) alpha and beta, Erythropoietin(***EPO***), Endothelial PAS Domain Protein 1(***EPAS1***), Egl nine homolog 1(***EGLN1***), Angiotensinogen (***AGT***), Angiotensin-Converting Enzyme (***ACE***), Endothelin 1 (***ET-1***), 70 kilodalton heat shock protein (***Hsp70***). We strongly believe that these proteins play a significant role in high-altitude diseases and can act as potential biomarkers.

The most frequently occurring event types—gene expression, regulation, binding, and protein catabolism—were further analyzed for their roles in HAD-related pathways and biomarker discovery. Notably, *gene expression* was linked to conditions such as chronic mountain sickness (CMS) and high-altitude pulmonary hypertension (HAPH). Genes like HIF, ROS-AMP, and ROE were heavily involved in these events and are known to regulate energy metabolism, catecholamine production, iron homeostasis, vasomotor control, and angiogenesis—key processes in cellular oxygen adaptation. HIF genes, in particular, function as transcriptional regulators and were originally identified in hepatoma cells as inducers of erythropoietin (EPO), which plays a protective role against ischemia in CMS.

In *binding events*, proteins such as EPAS1, EGLN1, and HSP70 were actively involved, revealing their functional roles in the molecular pathways associated with acute mountain sickness (AMS). Meanwhile, regulatory events involving HIF-1α, KOF-2, HSP90, and IL-8 highlighted their significance in controlling hypoxia-related gene activity. Lastly, *protein catabolism* events featured genes like HIF, HIF-1α, and EPO, which are potentially implicated in the gastrointestinal complications observed in altitude-related illnesses. The overall distribution and frequency of these biomolecular events across HAD literature are illustrated in Figure 3, offering insight into their potential as drug targets or diagnostic biomarkers.

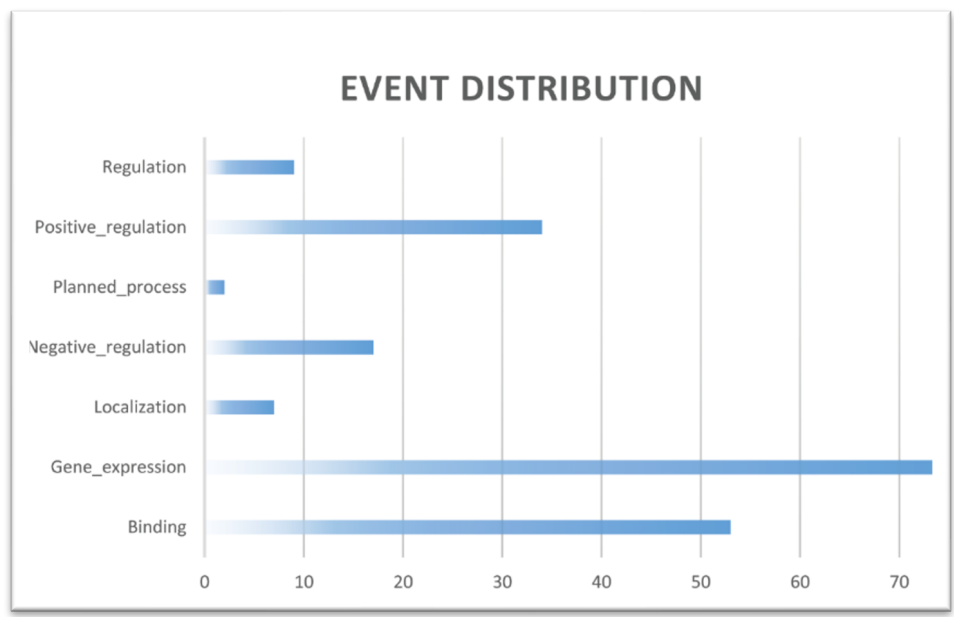

**Figure 3.** Distribution of biomolecular events in High Altitude Diseases

As shown in Figure 3, *gene expression*, *binding*, and *positive regulation* are the most frequently occurring biomolecular events associated with High-Altitude Diseases (HAD). To prioritize key molecular players within this event-driven network, we applied the PageRank algorithm, a widely adopted method in network analysis. PageRank assigns a numerical weight to each node—such as a gene or protein—based on its connectivity and influence within the network. This measure reflects not only the number of connections a node has but also the significance of the nodes it is connected to.

The PageRank analysis revealed several high-impact genes, including EPO (0.0163), VEGF (0.0148), HIF-1 (0.0136), EPAS1 and ACE (0.0119), and EGLN1, HSP70, ET-1 (0.0118). These proteins emerged as central entities within the network due to their frequent and diverse involvement in key events. Their prominent ranks underscore their functional relevance in hypoxia adaptation, vascular regulation, and erythropoiesis—critical biological responses in HAD. The full results of the PageRank analysis, including numerical rankings and connectivity scores, are provided in Supplementary Data 4, and the full biomolecular event network file is available as Supplementary Data 5.

By leveraging the PageRank algorithm, we identified *gene expression*, *regulation*, *binding*, and *protein catabolism* as event hubs that orchestrate the molecular pathways involved in HAD. The corresponding high-ranking proteins—EPO, VEGF, HIF-1, HSP70, ACE, EGLN1, and ET-1—can be considered potential biomarkers or therapeutic targets. Their network attributes, such as degree, proximity, and normalized min-max rank, are summarized in Figure 4, while Figure 5 presents the distribution of key network metrics, further validating their central roles in the HAD biomolecular landscape.

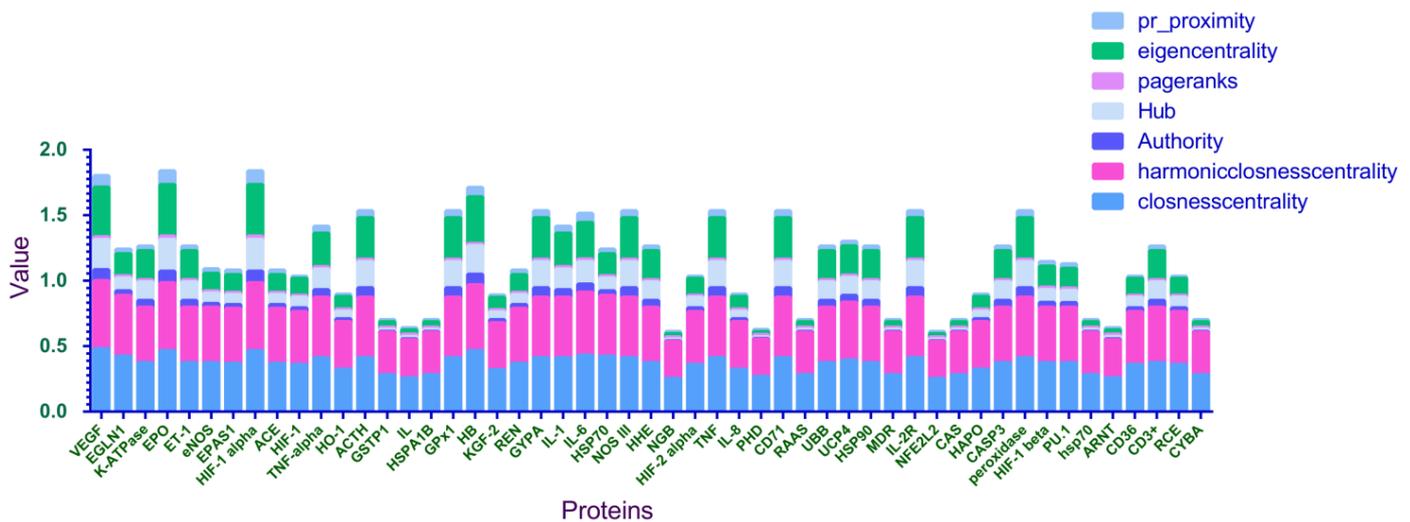

**Figure 4.** Top genes/proteins in biomolecular events of High Altitude Diseases based on the degree and PageRank within the network

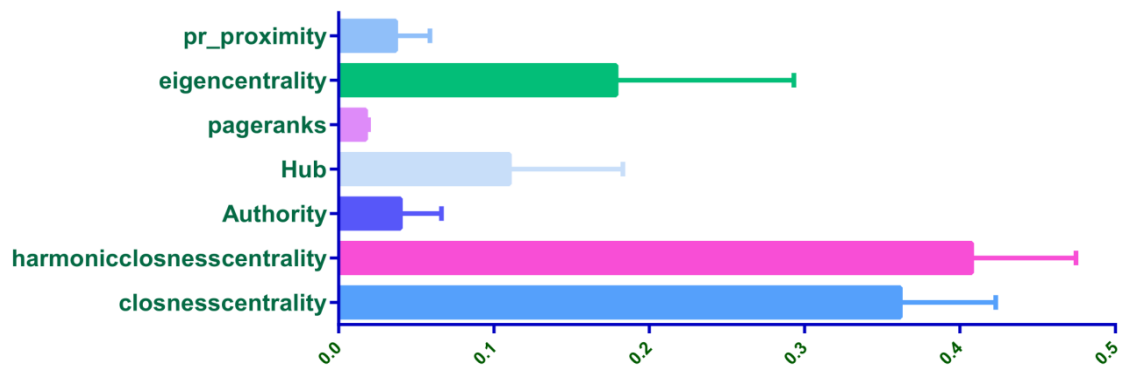

**Figure 5.** Distribution of the various network statistical measures applied to the biomolecular event network of High Altitude Diseases

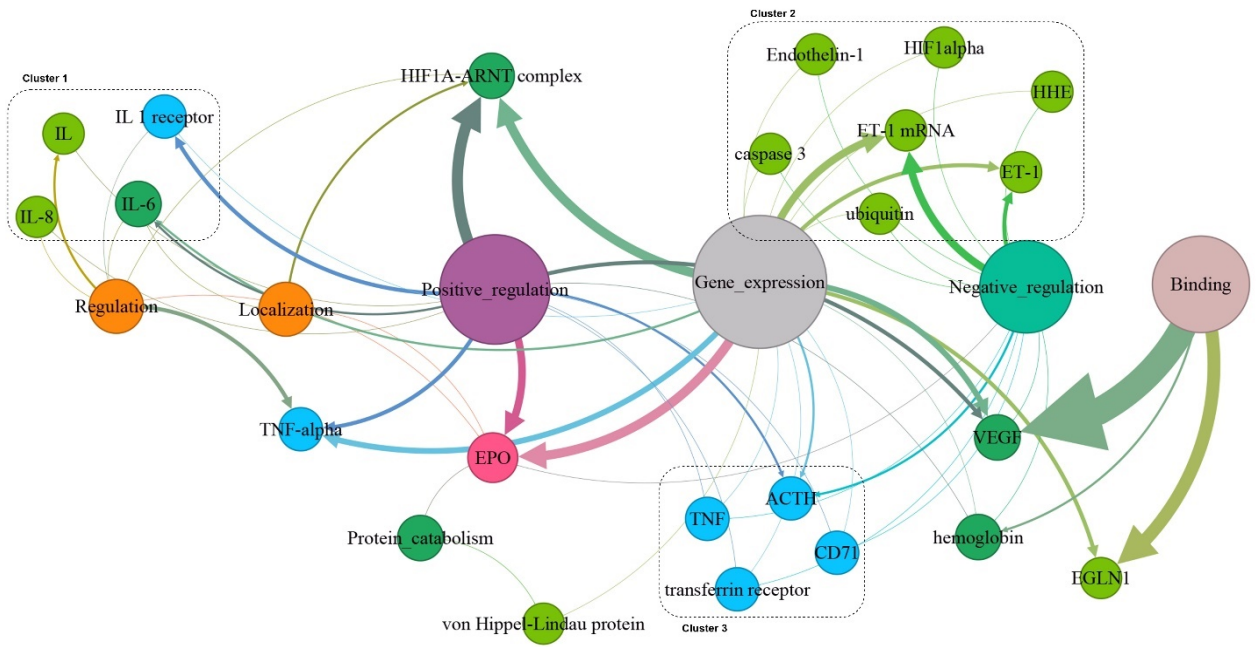

**Figure 6.** Biomolecular event subnetwork for high-altitude disease illustrating key molecular entities and biological processes involved in the response to hypoxic conditions.

Figure 6 shows the subnetwork of complex molecular interactions underlying high-altitude disease, emphasizing hypoxia-driven regulatory pathways. The network is organized into three major clusters: Cluster 1 encompasses inflammatory cytokines such as IL-6 and IL-8, which modulate immune and localization responses; Cluster 2 centers on hypoxia-inducible genes like HIF1α, ET-1, and endothelin-1, which contribute to vascular adaptation through negative gene regulation; and Cluster 3 involves stress-related and iron-regulating factors such as ACTH, TNF, and the transferrin receptor. Key processes such as positive regulation and gene expression serve as central hubs linking hypoxia signaling (e.g., HIF1A-ARNT complex) to downstream effectors like EPO and VEGF, highlighting their role in erythropoiesis and vascular remodeling. This systems-level representation underscores the interplay of immune, endocrine, and transcriptional responses critical to physiological adaptation and pathology at high altitude. We searched for biological databases for the major genes/proteins and the connected biological pathways, and provided substantial evidence in Table 2.

Table 2 summarizes the major genes and proteins identified through biomolecular event extraction and network analysis, along with their associated biological pathways relevant to high-altitude diseases (HAD). Key pathways include the HIF-1 signaling pathway, which features central hypoxia-responsive genes such as EPO, VEGFA, HIF1A, EPAS1, and EGLN1, all of which are critical for oxygen sensing, erythropoiesis, and vascular remodeling under hypoxic conditions. The renin-angiotensin system (RAS) pathway includes AGT, ACE, and AGTR1, implicating blood pressure regulation in HAD pathophysiology. The vascular smooth muscle contraction pathway involves genes like MAPK1 and EDNRA, indicating roles in vasoconstriction and vascular tone. Additionally, genes such as IL6, IL8, and TNF are involved in inflammatory and immune responses, while HSP70 and HSP90 represent key stress-related proteins contributing to cellular protection mechanisms. This comprehensive mapping of genes to pathways highlights the multifactorial nature of HAD and underscores the potential of these biomolecules as diagnostic or therapeutic targets.

**Table 2** Major Genes/Proteins and the Connected Biological Pathways in High-Altitude Diseases

| Pathway | Genes/Proteins | Biological Database |
|---|---|---|
| HIF-1 signaling pathway *(hsa04066)* | EPO (erythropoietin), VEGFA (vascular endothelial growth factor A), HIF1A (hypoxia-inducible factor 1 alpha), EPAS1, LDHA, NOS3, PHD1, EDN1, AKT3, ANGPT1 | KEGG Pathway<br>Reactome (e.g., "Signaling by HIF")<br>WikiPathways (e.g., WP1249 "HIF-1 signaling")<br>PANTHER Pathways |
| Renin-angiotensin system (RAS) *(hsa04614)* | AGT (angiotensinogen), AGTR1, AGTR2, ACE (angiotensin-converting enzyme), REN | KEGG Pathway<br>Reactome (e.g., "Renin-angiotensin system")<br>WikiPathways (e.g., WP1801 "Renin-Angiotensin System")<br>PANTHER Pathways |
| Vascular smooth muscle contraction *(hsa04270)* | AGTR1, EDNRA (endothelin receptor type A), MAPK1 | KEGG Pathway<br>Reactome (e.g., portions under "Smooth muscle contraction")<br>WikiPathways (e.g., WP1384 "Vascular smooth muscle contraction")<br>PANTHER Pathways |
| Hypoxia response & iron regulation | EGLN1, HIF1A, EPAS1, TFRC (transferrin receptor), LDHA | Reactome "Cellular response to hypoxia" "Iron uptake and transport"<br>WikiPathways (e.g., WP1445 "Hypoxia response"; WP133 "Iron metabolism")<br>Pathway Commons<br>PANTHER (under "Hypoxia response" and "Iron homeostasis") |
| Inflammatory signaling & localization | IL6, IL8, TNF, ACTH | Reactome (e.g., "Cytokine Signaling in Immune system")<br>KEGG Pathway (e.g., "Cytokine–cytokine receptor interaction" hsa04060)<br>WikiPathways (e.g., WP492 "Cytokine signaling in immune response")<br>PANTHER Pathways ("Inflammation mediated by chemokine and cytokine signaling pathway") |
| Protein folding & stress response | HSP70, HSP90 | Reactome (e.g., "Chaperonin-mediated protein folding"; "HSF1 activation")<br>WikiPathways (e.g., WP2806 "Heat shock response")<br>PANTHER Pathways ("Heat shock response")<br>KEGG Pathway ("Protein processing in endoplasmic reticulum" hsa04141 includes chaperones) |

Finally, we investigated the functional roles of some of these genes and identified that EGLN1 is an oxygen sensor catalyzer, EPAS1 plays a role in the rise to the blood-brain barrier, VEGFA induces the blood vessels permeabilization, and AGT is a potent blood pressure regulator [49-51]. Since HAD has been related to a lack of oxygen and hypertension, the roles of these genes are vital in the exploration of future studies. Major genes/proteins, their biological function, and the connected biological pathways in high-altitude diseases are given in Supplementary Data 6.

**Discussion**

High-altitude diseases (HAD), including AMS, HACE, and HAPE, represent a multifactorial physiological challenge involving oxygen deprivation, vascular dysregulation, inflammation, and stress response. Despite the growing number of clinical and molecular studies, there remains a critical gap in systematically identifying key biomolecules and pathways associated with HAD. In this study, we demonstrated the value of biomedical text mining as a scalable strategy to extract biomolecular events from literature, enabling a data-driven approach to biomarker discovery. By integrating named entity recognition, syntactic parsing, and a hybrid ensemble classification model, we identified high-confidence biological events such as gene expression, regulation, binding, and localization from over 7,800 PubMed abstracts related to HAD. These events offer a granular representation of how genes and proteins interact during hypoxic stress, and when organized into a network structure, they offer deeper insight into mechanistic relationships. PageRank-based analysis added an interpretive layer, ranking biomolecules by their connectivity and influence within the event network, thus aiding in prioritizing biomarker candidates for further investigation.

The biomolecular event network (Figure 2) constructed from the extracted events consisted of 97 nodes and 153 edges, where nodes represent either biomolecular entities or events, and edges capture co-occurrence relationships derived from the literature. Events like gene expression, binding, positive/negative regulation, and protein catabolism emerged as central hubs, reflecting their frequent and functionally diverse involvement in HAD. Proteins with the highest PageRank scores—such as EPO, VEGF, HIF-1α, EPAS1, EGLN1, ACE, ET-1, and HSP70—demonstrated widespread connectivity across multiple event types. This suggests their systemic involvement in physiological adaptations to hypoxia, including erythropoiesis, angiogenesis, and cellular stress response. Notably, the repeated association of these proteins with multiple event types (as both Theme and Cause) strengthens their biological relevance and implies their potential utility as biomarkers or therapeutic targets. For instance, the prominent role of HIF-1[52-54] and its downstream targets like EPO [55] and VEGF [56] aligns well with established knowledge on hypoxia signaling and high-altitude adaptation, thereby validating the model's ability to recover biologically significant relationships from unstructured text. To further refine our understanding, we performed a subnetwork analysis (Figure 6) that illuminated clustered patterns of molecular interaction within the main network. This analysis revealed three functional clusters: the first featured inflammatory mediators such as IL-6 [57] and IL-8 [58-59], which are known to modulate immune activation and vascular permeability in high-altitude exposure; the second was centered on hypoxia-inducible regulators like HIF1α [52-54], EGLN1[60-62], and EPAS1[60-62], reflecting their established role in vascular homeostasis and oxygen sensing; and the third included stress-related and metabolic regulators such as ACTH [63-65], TNF[63-65], and transferrin receptor, highlighting links to neuroendocrine signaling and iron metabolism. Central hubs in the subnetwork—positive regulation, binding, and gene expression—linked these clusters, showing how a relatively small set of recurring event types mediate diverse physiological processes in HAD. The subnetwork layout also highlighted the bidirectional and multi-event participation of top-ranked genes, underscoring their integrative role across pathways. These findings reinforce the utility of subnetwork analysis in dissecting complex molecular relationships and offer a focused map of high-value biomarker targets for future experimental validation.

This study demonstrates the effectiveness of combining biomedical text mining with network-based analysis to uncover key biomolecular players and regulatory mechanisms in high-altitude diseases. By systematically extracting and analyzing biomolecular events from literature, we constructed a comprehensive event network and identified central hub proteins using the PageRank algorithm. These high-confidence proteins—such as EPO, VEGF, HIF-1α, EPAS1, EGLN1, ACE, ACTH, TNF, ET-1, IL-6, IL-8 and HSP70—are strongly implicated in hypoxia adaptation and offer promising leads for further investigation as biomarkers or therapeutic targets. Moving forward, the integration of additional data sources such as gene expression profiles from high-altitude exposed populations, proteomics, or metabolomics datasets could significantly enhance the network's biological richness and predictive power. Incorporating such multi-omics evidence into the existing literature-derived network will enable more precise identification of disease modules, dynamic signaling cascades, and context-specific biomarkers, ultimately facilitating translational research and personalized intervention strategies in altitude-related pathophysiology.

**Conclusion**

Text mining offers powerful means to uncover hidden associations and novel patterns within the biomedical literature. In this study, we built a specialized corpus of 7,847 PubMed records to identify key genes and proteins involved in High Altitude Diseases (HAD). By applying the BCC-NER tool for gene/protein recognition and a hybrid machine-learning classifier (combining feature-based and MLG kernels) for event extraction, we assembled a biomolecular interaction network and used PageRank to highlight the most influential nodes. Our analysis revealed that hypoxia-related events—particularly gene expression, binding, regulation, and localization—occur repeatedly in the context of HAD. Central genes such as VEGFA, HIF1A, EPO, EPAS1, and EGLN1 emerged as critical hubs in hypoxia signaling, while AGT, ACE, and ET-1 connected hypoxic responses to blood-pressure regulation. HSP70 also featured prominently, reflecting its role in stress adaptation under low-oxygen conditions. Together, these biomolecular events and their associated genes underscore the intricate cross-talk between oxygen sensing, vascular homeostasis, and stress response pathways in HAD. Further investigation of these hypoxia-driven events is likely to yield valuable biomarkers and therapeutic targets for mitigating the physiological challenges of high-altitude exposure.

**Conflicts of interest**

The authors declare no conflicts of interest.

**Acknowledgments**

This work was supported by DRDO- BU Centre for Life Sciences, Coimbatore, Tamilnadu, India. BB acknowledges the fellowship received from the grant.

**Supplementary Data**

1. Identified bio-molecular events with PMID and corresponding abstracts
2. Identified Protein names with Theme ID
3. Identified Events with Theme and Event ID
4. Page Rank Information
5. Major genes/proteins and its biological function in High altitude diseases

**References**

[1] Gatterer, H., Villafuerte, F. C., Ulrich, S., Bhandari, S. S., Keyes, L. E., & Burtscher, M. (2024). Altitude illnesses. *Nature Reviews Disease Primers*, *10*(1), 43.
[2] Richalet, J. P., Hermand, E., & Lhuissier, F. J. (2024). Cardiovascular physiology and pathophysiology at high altitude. *Nature Reviews Cardiology*, *21*(2), 75-88.
[3] Li, X., Zhang, J., Liu, G., Wu, G., Wang, R., & Zhang, J. (2024). High altitude hypoxia and oxidative stress: the new hope brought by free radical scavengers. *Life Sciences*, *336*, 122319.
[4] Walsh, B., & Agrawal, S. (2024). High-altitude Cerebral Edema and High-altitude Pulmonary Edema Diagnosed in the Desert: A Case Report. *Clinical Practice and Cases in Emergency Medicine*, *8*(3), 202.
[5] Bhasuran, Balu. "Combining Literature Mining and Machine Learning for Predicting Biomedical Discoveries." Biomedical Text Mining. New York, NY: Springer US, 2022. 123-140.
[6] Bhasuran, Balu. "BioBERT and Similar Approaches for Relation Extraction." Biomedical Text Mining. New York, NY: Springer US, 2022. 221-235.
[7] Natarajan, Jeyakumar, Balu Bhasuran, and Gurusamy Murugesan. "Big Data Analytics: A Text Mining Perspective and Applications in Biomedicine and Healthcare." *Big Data Applications in Industry 4.0*. Auerbach Publications, 2022. 367-408.
[8] Bhasuran, Balu, and Jeyakumar Natarajan. "DisGeReExT: a knowledge discovery system for exploration of disease–gene associations through large-scale literature-wide analysis study." Knowledge and Information Systems (2023): 1-25.
[9] Silverman, A. L., Bhasuran, B., Mosenia, A., Yasini, F., Ramasamy, G., Banerjee, I., ... & Rudrapatna, V. A. (2025). Accurate, robust, and scalable machine abstraction of mayo endoscopic subscores from colonoscopy reports. *Inflammatory Bowel Diseases*, *31*(3), 665-670.

[10] Bhasuran, Balu, Gurusamy Murugesan, and Jeyakumar Natarajan. "Literature Based Discovery (LBD): Towards Hypothesis Generation and Knowledge Discovery in Biomedical Text Mining." arXiv preprint arXiv:2310.03766 (2023)

[11] Bhasuran B, Natarajan J. Distant supervision for large-scale extraction of gene–disease associations from literature using DeepDive. In: Lecture Notes in Networks and Systems. ; 2019. doi:10.1007/978-981-13-2354-6_39

[12] Bhasuran B, Subramanian D, Natarajan J. Text mining and network analysis to find functional associations of genes in high altitude diseases. Comput Biol Chem. 2018. doi:10.1016/j.compbiolchem.2018.05.002

[13] Maroli, N., Kalagatur, N. K., Bhasuran, B., Jayakrishnan, A., Manoharan, R. R., Kolandaivel, P., ... & Kadirvelu, K. (2019). Molecular mechanism of T-2 toxin-induced cerebral edema by aquaporin-4 blocking and permeation. Journal of chemical information and modeling, 59(11), 4942-4958.

[14] Maroli, N., Bhasuran, B., Natarajan, J., & Kolandaivel, P. (2020). The potential role of procyanidin as a therapeutic agent against SARS-CoV-2: a text mining, molecular docking and molecular dynamics simulation approach. Journal of Biomolecular Structure and Dynamics, 1-16.

[15] Bhasuran, B., Schmolly, K., Kapoor, Y., Jayakumar, N. L., Doan, R., Amin, J., ... & Rudrapatna, V. A. (2025). Reducing diagnostic delays in acute hepatic porphyria using health records data and machine learning. *Journal of the American Medical Informatics Association*, *32*(1), 63-70.

[16] Wu, P., Li, X., Gu, J., Qian, L., & Zhou, G. (2024). Pipelined biomedical event extraction rivaling joint learning. *Methods*, *226*, 9-18.

[17] Liu, L., Liu, M., Liu, S., & Ding, K. (2024). Event extraction as machine reading comprehension with question-context bridging. *Knowledge-Based Systems*, *299*, 112041.

[18] Frisoni, G., Moro, G., & Carbonaro, A. (2021). A survey on event extraction for natural language understanding: Riding the biomedical literature wave. *IEEE Access*, *9*, 160721-160757.

[19] Fei, H., Ren, Y., Zhang, Y., Ji, D., & Liang, X. (2021). Enriching contextualized language model from knowledge graph for biomedical information extraction. *Briefings in bioinformatics*, *22*(3), bbaa110.

[20] van Landeghem S, Björne J, Wei CH, et al. Large-Scale Event Extraction from Literature with Multi-Level Gene Normalization. PLoS One. 2013. doi:10.1371/journal.pone.0055814

[21] McClosky D, Riedel S, Surdeanu M, McCallum A, Manning CD. Combining joint models for biomedical event extraction. BMC Bioinformatics. 2012. doi:10.1186/1471-2105-13-S11-S9

[22] Björne J, Salakoski T. Generalizing Biomedical Event Extraction. Proc BioNLP 2011 Work Companion Vol Shar Task Portl Oregon June Assoc Comput Linguist. 2011.

[23] Gerner M, Sarafraz F, Bergman CM, Nenadic G. BioContext: An integrated text mining system for large-scale extraction and contextualization of biomolecular events. Bioinformatics. 2012. doi:10.1093/bioinformatics/bts332

[24] Riedel S, Chun H-W, Takagi T, Tsujii J. A Markov logic approach to bio-molecular event extraction. In: ; 2010. doi:10.3115/1572340.1572347

[25] Zhou D, Zhong D, He Y. Biomedical Relation Extraction: From Binary to Complex. Comput Math Methods Med. 2014. doi:10.1155/2014/298473

[26] Ohta T, Pyysalo S, Tsujii J. From pathways to biomolecular events: opportunities and challenges. In: Proceedings of BioNLP 2011 Workshop. ; 2011:105-113.

[27] Hakala K, Landeghem S Van. EVEX in ST'13: Application of a large-scale text mining resource to event extraction and network construction. Acl 2013. 2013.

[28] Abdulkadhar, Sabenabanu, Balu Bhasuran, and Jeyakumar Natarajan. "Multiscale Laplacian graph kernel combined with lexico-syntactic patterns for biomedical event extraction from literature." Knowledge and Information Systems 63.1 (2021): 143-173.

[29] Kwartler T. The OpenNLP Project. In: Text Mining in Practice with R. ; 2017. doi:10.1002/9781119282105.ch8

[30] Liu H, Christiansen T, Baumgartner WA, Verspoor K. BioLemmatizer: A lemmatization tool for morphological processing of biomedical text. J Biomed Semantics. 2012. doi:10.1186/2041-1480-3-3

[31] Richard Socher and John Bauer and Christopher D. Manning and Andrew Y. Ng, Socher R, Bauer J, Manning CD, Ng AY. 9-1 Parsing With Compositional Vector Grammars. Proc {ACL} Conf. 2013.

[32] Sagae K, Tsujii J. Dependency Parsing and Domain Adaptation with Data-Driven LR Models and Parser Ensembles. In: ; 2010. doi:10.1007/978-90-481-9352-3_4

[33] Murugesan G, Abdulkadhar S, Bhasuran B, Natarajan J. BCC-NER: bidirectional, contextual clues named entity tagger for gene/protein mention recognition. Eurasip J Bioinforma Syst Biol. 2017. doi:10.1186/s13637-017-0060-6

[34] Kondor R, Pan H. The multiscale laplacian graph kernel. In: Advances in Neural Information Processing Systems. ; 2016:2990-2998.

[35] Claesen M, De Smet F, Suykens JAK, De Moor B. EnsembleSVM: A library for ensemble learning using support vector machines. *J Mach Learn Res*. 2014;15(1):141-145.

[36] Bhasuran B, Murugesan G, Abdulkadhar S, Natarajan J. Stacked ensemble combined with fuzzy matching for biomedical named entity recognition of diseases. J Biomed Inform. 2016;64. doi:10.1016/j.jbi.2016.09.009

[37] Bhasuran B, Natarajan J. Automatic extraction of gene-disease associations from literature using joint ensemble learning. PLoS One. 2018;13(7). doi:10.1371/journal.pone.0200699

[38] Bastian M, Heymann S, Jacomy M. Gephi: an open source software for exploring and manipulating networks. In: *Third International AAAI Conference on Weblogs and Social Media*. ; 2009.

[39] Page L. PageRank citation ranking: bring order to the web. *Stanford Digit Libr Work Pap*. 1997.

[40] Zhu Y, Song M, Yan E. Identifying liver cancer and its relations with diseases, drugs, and genes: A literature-based approach. *PLoS One*. 2016. doi:10.1371/journal.pone.0156091

[41]  Aftiss, A., Lamsiyah, S., El Alaoui, S. O., & Schommer, C. (2024). BioMDSum: An Effective Hybrid Biomedical Multi-Document Summarization Method Based on PageRank and Longformer Encoder-Decoder. IEEE Access.

[42]  Gupta, S., Sharaff, A., & Nagwani, N. K. (2023). Graph Ranked Clustering Based Biomedical Text Summarization Using Top k Similarity. Computer Systems Science & Engineering, 46(1).

[43]  Ma, K., Li, J., Zhao, M., Zamit, I., Lin, B., Guo, F., & Tang, J. (2024). PPRTGI: A Personalized PageRank Graph Neural Network for TF-Target Gene Interaction Detection. IEEE/ACM Transactions on Computational Biology and Bioinformatics.

[44]  Deng, J. (2025). Enhancing peer review in educational assessments via social media platforms: A Fuzzy-CDF and weighted PageRank approach. *Journal of Computational Methods in Sciences and Engineering*, *25*(2), 1303-1315.

[45]  Zhang, Z., Jia, Y., Zhang, X., Papadopoulou, M., & Roche, C. (2023). Weighted co-occurrence bio-term graph for unsupervised word sense disambiguation in the biomedical domain. *IEEE Access*, *11*, 45761-45773.

[46]  Lima, T., Ferreira, R., Freitas, M., Henrique, R., Vitorino, R., & Fardilha, M. (2022). Integration of automatic text mining and genomic and proteomic analysis to unravel prostate cancer biomarkers. *Journal of Proteome Research*, *21*(2), 447-458.

[47]  Li, X., Dai, A., Tran, R., & Wang, J. (2023). Identifying miRNA biomarkers for breast cancer and ovarian cancer: a text mining perspective. *Breast Cancer Research and Treatment*, *201*(1), 5-14.

[48]  Lever, J., Jones, M. R., Danos, A. M., Krysiak, K., Bonakdar, M., Grewal, J. K., ... & Jones, S. J. (2019). Text-mining clinically relevant cancer biomarkers for curation into the CIViC database. *Genome medicine*, *11*, 1-16.

[49]  Wang, S., Hong, Q., Zheng, Y., Duan, S., Cai, G., & Chen, X. (2024). Epigenetics in high-altitude adaptation and disease. *Science Bulletin*.

[50]  Chen Y, Jiang C, Luo Y, Liu F, Gao Y. Interaction of CARD14, SENP1 and VEGFA polymorphisms on susceptibility to high altitude polycythemia in the Han Chinese population at the Qinghai--Tibetan Plateau. *Blood Cells, Mol Dis*. 2016;57:13-22.

[51]  Bhattacharyya, P., Deb, P., & Pradhan, D. (2023). Genetics in High Altitude Medicine. In *High Altitude Medicine: A Case-Based Approach* (pp. 23-29). Cham: Springer International Publishing.

[52]  Yfantis, A., Mylonis, I., Chachami, G., Nikolaidis, M., Amoutzias, G. D., Paraskeva, E., & Simos, G. (2023). Transcriptional response to hypoxia: the role of HIF-1-associated co-regulators. *Cells*, *12*(5), 798.

[53]  Lyu, Y., Yang, Y., Talwar, V., Lu, H., Chen, C., Salman, S., ... & Semenza, G. L. (2024). Hypoxia-inducible factor 1 recruits FACT and RNF20/40 to mediate histone ubiquitination and transcriptional activation of target genes. *Cell Reports*, *43*(4).

[54]  Koyasu, S., Horita, S., Saito, K., Kobayashi, M., Ishikita, H., Chow, C. C., ... & Harada, H. (2023). ZBTB2 links p53 deficiency to HIF-1-mediated hypoxia signaling to promote cancer aggressiveness. *EMBO reports*, *24*(1), e54042.

[55]  Kourembanas S, Marsden PA, McQuillan LP, Faller DV. Hypoxia induces endothelin gene expression and secretion in cultured human endothelium. J Clin Invest. 1991 Sep;88(3):1054-7. doi: 10.1172/JCI115367. PMID: 1885767; PMCID: PMC295521.

[56] Droma, Y., Hanaoka, M., Kinjo, T., Kobayashi, N., Yasuo, M., Kitaguchi, Y., & Ota, M. (2022). The blunted vascular endothelial growth factor-A (VEGF-A) response to high-altitude hypoxia and genetic variants in the promoter region of the VEGFA gene in Sherpa highlanders. *PeerJ*, *10*, e13893.

[57] Hartmann G, Tschöp M, Fischer R, Bidlingmaier C, Riepl R, Tschöp K, Hautmann H, Endres S, Toepfer M. High altitude increases circulating interleukin-6, interleukin-1 receptor antagonist and C-reactive protein. Cytokine. 2000 Mar;12(3):246-52. doi: 10.1006/cyto.1999.0533. PMID: 10704252.

[58] Karakurum, M., Shreeniwas, R., Chen, J., Pinsky, D., Yan, S. D., Anderson, M., ... & Stern, D. (2021). Hypoxic induction of interleukin-8 gene expression in human endothelial cells. *The Journal of clinical investigation*, *93*(4), 1564-1570.

[59] Wang, Z. Y., Liu, Y., Li, S. P., Li, J. J., Zhang, Z., Xiao, X. C., ... & Yang, S. (2022). Hypoxia inducible factor 1α promotes interleukin-1 receptor antagonist expression during hepatic ischemia-reperfusion injury. *World Journal of Gastroenterology*, *28*(38), 5573.

[60] Gou, Q., & Xue, H. (2025). ASSOCIATION STUDY OF EPAS1 GENE POLYMORPHISMS WITH THE SUSCEPTIBILITY TO HIGH ALTITUDE PULMONARY HYPERTENSION IN TIBETAN RESIDENTS. *Journal of Hypertension*, *43*(Suppl 1), e203.

[61] Dai, Zhiyu. "Translational potential of hypoxia-inducible factor-2α signaling in pulmonary hypertension." Lung Biology and Pathophysiology (2024): 87-100.

[62] Julian, Colleen G., and Lorna G. Moore. "Human genetic adaptation to high altitude: evidence from the Andes." Genes 10.2 (2019): 150.

[63] Reid, Brie M., and Michael K. Georgieff. "The interaction between psychological stress and iron status on early-life neurodevelopmental outcomes." Nutrients 15.17 (2023): 3798.

[64] Zangeneh, Farideh Zafari, and Maryam Sarmast Shoushtari. "The Significant Role of the Sympathetic Nervous System in Hospitalization and Death of Patients With COVID-19: Psycho-Neuroendocrine-Immune Aspect of Stress." Acta Medica Iranica (2022).

[65] Huang, H., & Yang, L. (2023). Research Progress of Inflammatory Factors in Chronic Obstructive Pulmonary Disease with Pulmonary Hypertension at High Altitude. *Alternative Therapies in Health and Medicine*, *29*(8), 689-693.